\documentclass[12pt,preprint,longabstract]{aastex}
\usepackage[onecolumn,numberedappendix]{emulateapj5}

\begin{document}

\vspace{1cm}


\title{Dark Matter Annihilation and Primordial Star Formation}

\slugcomment{Submitted to ApJ}

\author{Aravind Natarajan}
\affil{Fakult\"{a}t f\"{u}r Physik, Universit\"{a}t Bielefeld, Universit\"{a}tsstra\ss e 25, Bielefeld D-33615, Germany}
\email{anatarajan@physik.uni-bielefeld.de}

\author{Jonathan C. Tan}
\affil{Department of Astronomy, University of Florida, Gainesville, FL 32611, U.S.A.}
\email{jt@astro.ufl.edu}

\author{Brian W. O'Shea}
\email{oshea@msu.edu}
\affil{Department of Physics \& Astronomy, Michigan State University, East Lansing, MI 48864, U.S.A.}

\begin{abstract}
We investigate the effects of weakly-interacting massive particle
(WIMP) dark matter annihilation on the formation of Population III.1
stars, which are theorized to form from the collapse of gas cores at
the centers of dark matter minihalos. We consider the relative
importance of cooling due to baryonic radiative processes and heating
due to WIMP annihilation. We analyze the dark matter and gas profiles
of several halos formed in cosmological-scale numerical
simulations. The heating rate depends sensitively on the dark matter
density profile, which we approximate with a power law
$\rho_{\chi}\propto r^{-\alpha_\chi}$, in the numerically
unresolved inner regions of the halo. If we assume a self-similar
structure so that $\alpha_{\chi}\simeq 1.5$ as measured on the
resolved scales $\sim 1$~pc, then for a fiducial WIMP mass of 100~GeV,
the heating rate is typically much smaller ($<10^{-3}$) than the
cooling rate for densities up to $n_{\rm H}=10^{17}\:{\rm
cm^{-3}}$. In one case, where $\alpha_{\chi}=1.65$, the heating rate
becomes similar to the cooling rate by a density of $n_{\rm
H}=10^{15}\:{\rm cm^{-3}}$. The dark matter density profile is
expected to steepen in the central baryon-dominated region $\lesssim
1$~pc due to adiabatic contraction, and we observe this effect (though
with relatively low resolution) in our numerical models. From these we
estimate $\alpha_{\chi}\simeq 2.0$. The heating now dominates cooling
above $n_{\rm H}\simeq 10^{14}\:{\rm cm^{-3}}$, in agreement with the
previous study of Spolyar, Freese \& Gondolo. We expect this leads to
the formation of an equilibrium structure with a baryonic and dark
matter density distribution exhibiting a flattened central
core. Examining such equilibria, we find total luminosities due to
WIMP annihilation are relatively constant and $\sim 10^3\:L_\odot$,
set by the radiative luminosity of the baryonic core. We discuss the
implications for Pop III.1 star formation, particularly the subsequent
growth and evolution of the protostar. Even if the initial protostar
fails to accumulate any additional dark matter, its contraction to the
main sequence could be significantly delayed by WIMP annihilation
heating, potentially raising the mass scale of Pop III.1 stars to
masses $\gg100\:M_\odot$.
\end{abstract}

\keywords{cosmology: theory --- dark matter --- early universe --- galaxies: formation --- stars: formation}

\maketitle

\section{Introduction}

Population III stars are defined to be those whose formation and
evolution are independent of metallicity (McKee \& Tan 2008; O'Shea et
al. 2008), since their metallicity is extremely low: close to or equal
to that arising from big bang nucleosynthesis. Population III.1 stars
are defined as having their formation be independent of other stars or
other astrophysical objects, so that their initial conditions are
determined solely by cosmology. These stars will be the first objects
to form in a given region of the universe and they are likely to play
an important role in cosmic reionization and in laying the foundations
for the formation and build-up of galaxies. It is possible that they
are the direct or indirect progenitors of supermassive black holes.

Within the commonly accepted $\Lambda$CDM framework, Pop III.1 stars
form within dark matter halos. Indeed, in those halos that do form
stars, only one star appears to form in the center of each halo (Abel
et al. 2002; Bromm, Coppi, \& Larson 2002).

One of the most theoretically well-motivated cold dark matter candidates is
a Weakly-Interacting Massive Particle (WIMP). Supersymmetric theories
with R-parity naturally provide a stable dark matter candidate which
could compose all or part of the dark matter in the Universe.

It has been pointed out that, if dark matter consists of a WIMP such
as the supersymmetric neutralino, the energy released by the
annihilation of these particles could influence early structure
formation, star formation and protostellar evolution (Ripamonti,
Mapelli \& Ferrara 2007; Ascasibar 2007; Spolyar, Freese, \& Gondolo
2008; Iocco 2008; Freese, Spolyar \& Aguirre 2008; Freese et
al. 2008b,c). The effects on stellar evolution at fixed mass have also
been investigated (Taoso et al. 2008; Yoon, Iocco \& Akiyama 2008).
Spolyar et al. (2008) show that, for the adiabatically-contracted
Navarro, Frenk \& White (1996) (NFW) dark matter density profiles they
considered, dark matter heating can overwhelm gas cooling in the
innermost region of a star-forming minihalo, and they propose that
this can then lead to a dark matter powered star.

In this paper we revisit the scenario investigated by Spolyar et
al. (2008). In \S\ref{S:analytic} we derive an analytic expression for
the dark matter heating rate, including a simplified treatment of
radiative transport. In \S\ref{S:results} we present our results of
the assessment of the importance of WIMP annihilation heating for
several halos formed in numerical simulations of cosmic structure
formation (O'Shea \& Norman 2007). We describe the dark matter density
structure in \S\ref{S:dm} and the properties of the baryons in
\S\ref{S:baryon}. We compare the WIMP annihilation heating rates and
the baryonic cooling rates and discuss the equilibrium structure of
dark matter powered protostars in \S\ref{S:equilibrium}. We discuss
the implications for subsequent protostellar evolution in
\S\ref{S:protostar}. We conclude in \S\ref{S:conclusions}.


\section{Analytic Prescription for WIMP Annihilation Heating}\label{S:analytic}

Let us consider WIMPs with a mass $m_\chi$. Let $<\sigma_a v>$ be the
WIMP annihilation cross section times the relative velocity, averaged
over the momentum distribution of the WIMPs. For numerical evaluations
in this paper we will assume it has a value $3 \times 10^{-26}\:{\rm
cm^3\:s^{-1}}$ (e.g. Jungman, Kamionkowski \& Griest 1996). The number
of photons produced by WIMP annihilation per unit volume, per unit
time and per unit energy, at the point $r'$ is (Hall \& Gondolo 2006;
Bergstr\"om, Ullio \& Buckley 2006):
\begin{equation}
\frac{d{\cal N}_{\gamma}}{dE_\gamma} = \frac{\rho^2_\chi (r') <\sigma_a v> }{2 m^2_\chi} \; \frac{dN_{\gamma}}{dE_\gamma},
\end{equation}
where $\rho_\chi (r')$ is the WIMP density at $r'$ and
$dN_{\gamma}/dE_\gamma$ is the number of photons produced per unit
energy per annihilation.

Let us consider electron scattering at a location $r$, by photons
produced at the location $r'$. Some of the photons produced at
$r'$ will scatter off electrons before reaching the location
$r$. The number of photons reaching $r$ without scattering is smaller
than the number produced at $r'$ by the factor $S$:
\begin{eqnarray}
S(r',r,E,\theta) = 
  \left \{     
    \begin{array}{ll}
    \exp \left[- \sigma_{e^-\gamma}(E_\gamma) \int_0^{s_+} ds \; n_e(s) \right ]        
             \mbox { $\;\;\;\;\;$ when $r' < r$ } \\
\exp \left[- \sigma_{e^-\gamma}(E_\gamma) \int_0^{s_-} ds \; n_e(s) \right ]\\ \times \left( 1 + \exp \left[- \sigma_{e^-\gamma}(E_\gamma)\int_{s_-}^{s_+} ds \; n_e(s) \right ] \right )     
        \mbox { when $r{'} > r$  }
    \end{array}
\right.  \label{S}
\end{eqnarray}
where $s_\pm = r' \cos\theta \pm \sqrt{ r^2 - r'^2 \sin^2 \theta
}$ and $\sigma_{e^-\gamma}$ is the cross section. Eq. \ref{S} takes
account of the fact that photons can pass through the sphere
$r=$constant twice when $r' > r$. We note that $S$ is a function of
the angle $\theta$ between the path of the photon and the line joining
$r'$ with the center of the spherical cloud.

Consider photons that travel from $r'$ to $r$. The number of
scattering events per unit energy, per unit time, per unit volume at
$r$, along the path $\theta$ is then
\begin{equation}
\frac{d{\cal N}_{s}}{dE_\gamma} = \frac{<\sigma_a v>}{2 m^2_\chi} \frac{dN_\gamma}{dE_\gamma}  \int dr' r'^2 \rho^2_\chi (r') \; S(r',r,E,\theta) \frac{ n_e(r) \sigma_{e^-\gamma}(E_\gamma) \; \delta s(r',r,\theta) }{  r^2 \; \delta r },
\end{equation}
where $\delta s$ is the distance between $r$ and $r+\delta r$, along the path of the photon:
\begin{equation}
\delta s = \frac{\delta r}{\sqrt{1 - \frac{r'^2}{r^2} \sin^2\theta}}.
\end{equation}
Since all angles $\theta$ are equally probable, we compute the angle averaged quantity $<\bar{S}>$:
\begin{equation}
<\bar{S}(r',r,E)> =  \frac{1}{2} \int_0^{\theta_{max}} d\theta \; \sin\theta \; \frac{S(r',r,E,\theta)}{\sqrt{1-\frac{r'^2}{r^2} \sin^2 \theta}}.
\label{S_avg}
\end{equation}
$\theta_{max} = \pi$ when $r' < r$ and $\theta_{max}=\sin^{-1}(r/r')$ when $r' > r$.

In the limit that the wavelength of the incident radiation is much
smaller than the Bohr radius, we may assume that the radiation sees a
gas of electrons and protons, rather than bound atoms. With this
approximation, the scattering cross section
$\sigma_{e^-\gamma}(E_\gamma)$ is given by the Klein-Nishina formula
(e.g. Peskin \& Schroeder 1995)
\begin{equation}
\sigma_{e^{-}\gamma} (y) = \sigma_T \; f_\sigma(y),
\end{equation}
where $\sigma_T$ is the Thomson cross section for electrons (proton
scattering is less important owing to the large mass). $f_\sigma$ is
given by
\begin{equation}
f_\sigma(y) = \frac{3}{8}  \left[ \frac{2(1+y)}{(1+2y)^2} + \frac{\ln(1 + 2y)}{y} - \frac{2}{1+2y} + \frac{2(1+y)^2}{y^2(1+2y)} - \frac{2(1+y) \ln(1+2y)}{y^3} + \frac{2}{y^2} \right],
\end{equation}
where $y = E_\gamma / m_e$. $E_{\rm abs}$ is the average energy
transferred to an electron in a scattering event:
\begin{equation}
E_{\rm abs} = m_\chi f_E(E_\gamma),
\end{equation}
where
\begin{equation}
f_E = \frac{E_\gamma}{m_\chi} + \frac{m_e}{m_\chi} - \frac{m_e}{2 m_\chi} \ln \left( 1 + \frac{2E_\gamma}{m_e} \right).
\end{equation}
Let us make the assumption that the energetic electrons (after
scattering) remain in the gas cloud and serve to heat up the cloud. As
long as low energy photons do not contribute significantly to the
heating rate, we may maintain our approximation $E_\gamma \gg 13.6$
eV, while performing the energy integral:
\begin{equation}
H(r) = \frac{<\sigma_a v> n_e(r) \sigma_T}{2 m_\chi r^2} \int_0^{m_\chi} dE_\gamma \;f_E(E_\gamma) f_\sigma(E_\gamma) \frac{dN_\gamma}{dE_\gamma}(E_\gamma) \int_0^r dr' r'^2 \rho^2_\chi(r') \left< \bar{S}(E_\gamma,r',r) \right >.
\label{heating}
\end{equation}
$H(r)$ gives the heating rate (energy per unit volume per unit time)
at the point $r$ due to primary photons, under the approximate
assumption that all the primary photon energy is transferred to the
gas at the location of the first collision. Note, we are not solving
the full radiation transfer problem in determining $H(r)$, especially
the distributed heating effects of secondary pair and photon
production, and so our treatment is still an approximate one.

\subsection { The $ \left < S (r,r',E) \right >$ function. }

\begin{figure}[!h]
\plotone{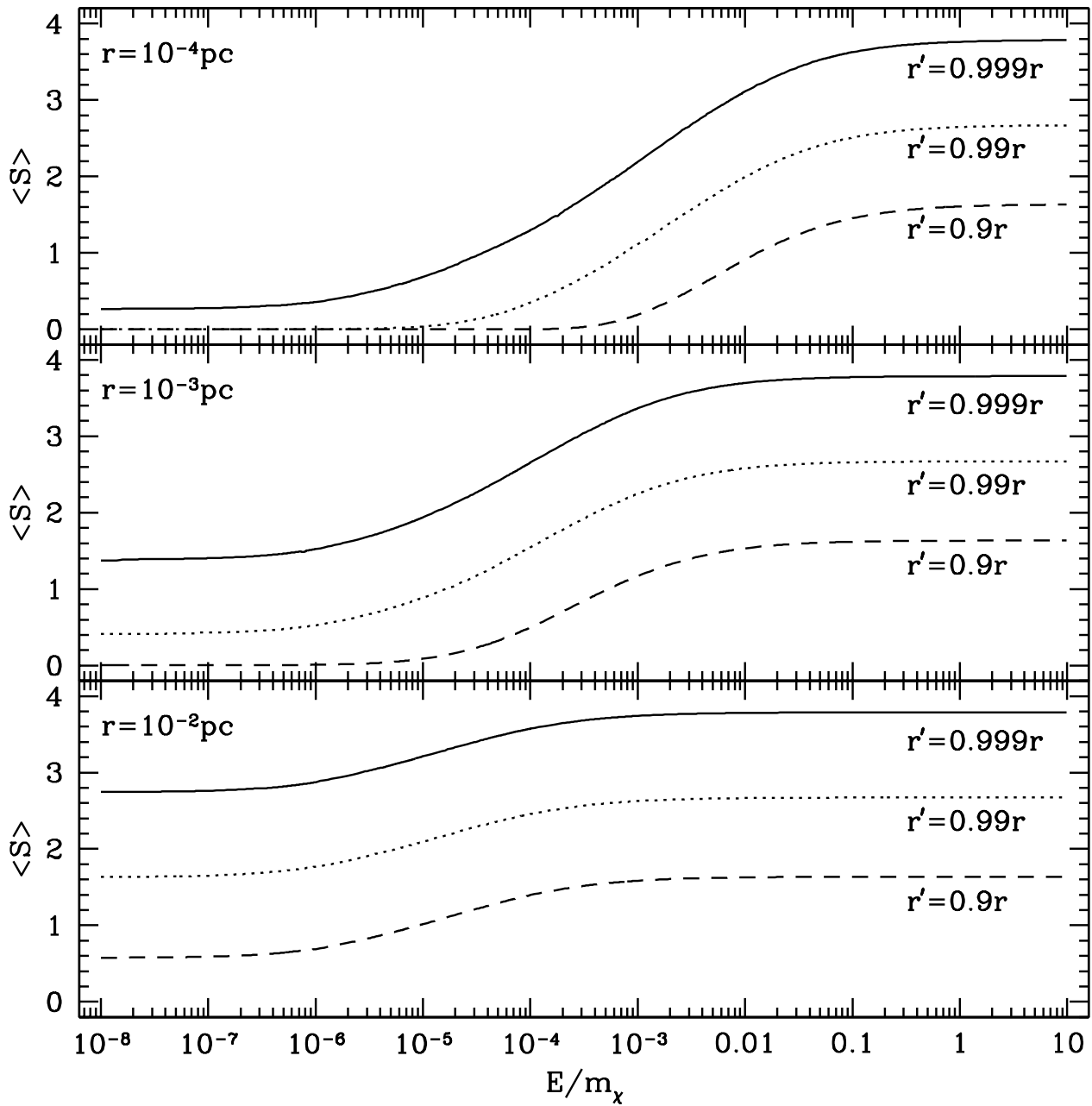}
\caption{The $\left<S(r,r',E)\right>$ function for $m_\chi=100$~GeV. $S$ is small for
small energies because of the larger scattering cross section. $S$ is
also small when the point $r'$ is far from $r$. These effects are
more pronounced for small $r$ (high gas density). \label{S_fig}}
\end{figure}

In order to compute the heating rate at a point $r$ using
Eq. \ref{heating}, we need to consider dark matter in a finite volume
around $r$. The size of this volume is determined by the function $
\left < S (r, r',E) \right >$. When this volume is small compared
to the size of the halo, one may expect most of the energy released by
dark matter annihilation to be absorbed by the gas
cloud. Figure~\ref{S_fig} shows $<S>$ for the cases $r = 10^{-4},
10^{-3}$ and $10^{-2}$ pc. The three curves in each panel correspond
to values of $r' = 0.999 r$, $r' = 0.99 r$ and $r' =
0.9r$. As expected, $S$ is largest when $r'$ is close to $r$. More
energetic photons penetrate farther through the gas owing to the
smaller scattering cross section.
   
\section{Numerical Results for WIMP Annihilation Heating and Baryonic Cooling Rates}\label{S:results}

We analyzed the structure of the three highest resolution star-forming
minihalos in the numerical simulations presented by O'Shea \& Norman
(2007). We will refer to these simulation runs as A, B, C. These
simulations were performed using the Enzo code, which is an adaptive
mesh refinement (AMR) cosmology code developed by G. Bryan and others
(e.g. O'Shea et al. 2004). The code couples an N-body particle mesh
(PM) solver (Efstathiou et al. 1985; Hockney \& Eastwood 1998) with
Eulerian AMR.  All simulations were initialized at $z = 99$ assuming a
concordance cosmological model with $\Omega_m = 0.3, \Omega_b = 0.04,
\Omega_{\rm CDM} = 0.26, \Omega_\Lambda = 0.7, h = 0.7, \sigma_8 = 0.9$
and an Eisenstein \& Hu (1999) power spectrum with a spectral index $n
= 1$. Each of the three star-forming minihalos was formed in a
separate simulation with a different random seeding of the initial
conditions and a box size of $0.3 h^{-1}$ Mpc (comoving). The most
massive halo to form in each simulation, at $z = 15$ (typically with a
mass $\sim 10^6 M_\odot$) was found using a dark matter only
computation. The initial conditions were then regenerated with both
dark matter and baryons such that the Lagrangian volume in which the
halo formed was resolved at much higher spatial and mass
resolution. 

\subsection{Dark Matter Density Structure and Heating Rate}\label{S:dm}

\begin{figure}[!h]
\plotone{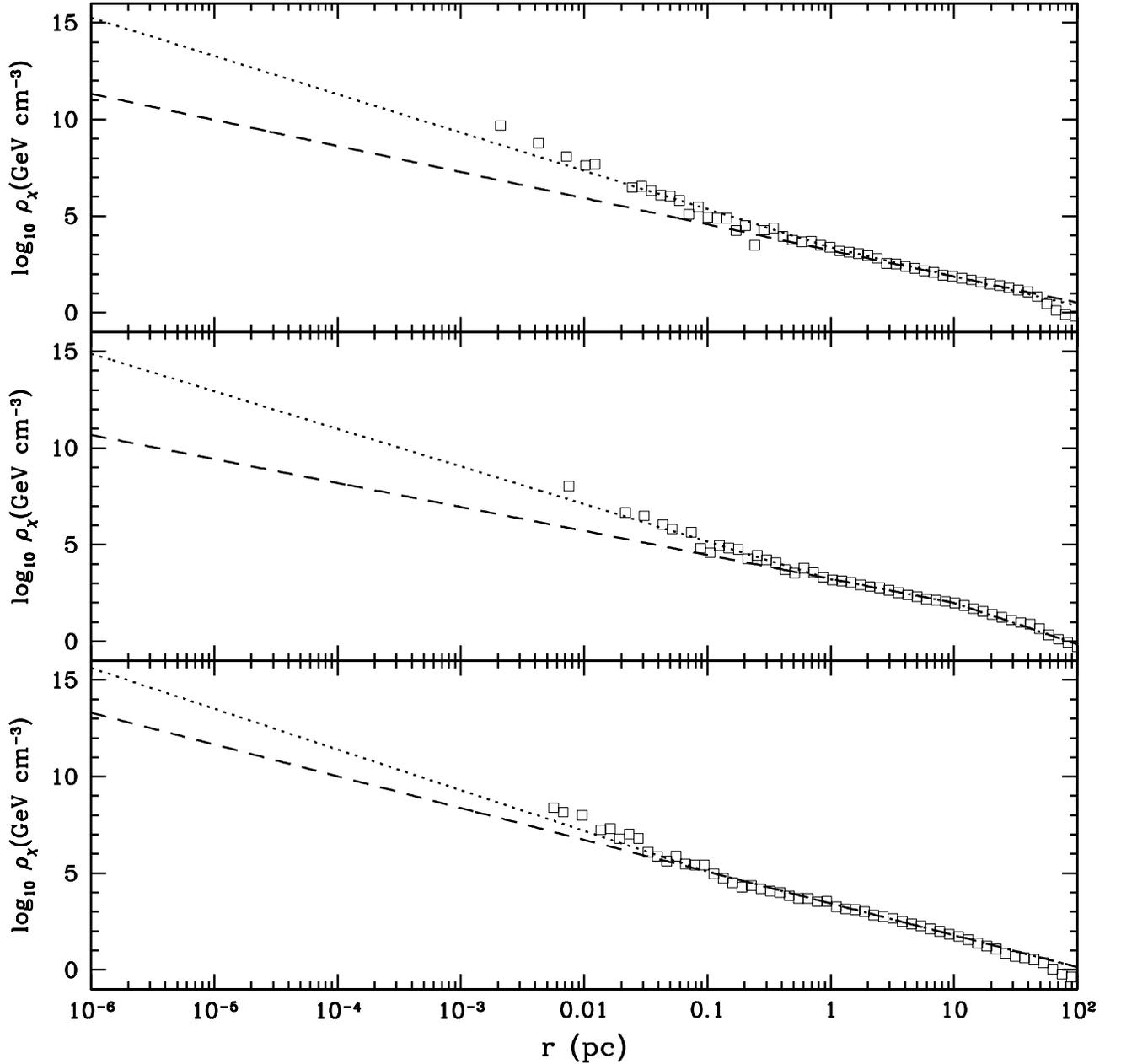}
\caption{Dark matter density profiles (spherical averages) for
simulations A (top), B (middle), and C (bottom) are shown with the
open squares. Power law fits to the outer, well-resolved regions (fit
\#1) are shown with the dashed lines, while fits to the inner (but not innermost -- see text),
less-well-resolved, regions (fit \#2) are shown with the dotted lines.
\label{dm}}
\end{figure}

The minimum dark matter particle mass in the simulations was $1.81
h^{-1}\:M_\odot \rightarrow2.59\:M_\odot$. To avoid effects due to the
finite size of the dark matter particles, the dark matter density was
smoothed on a comoving scale of 0.5 pc, i.e. a proper scale of $\sim
0.025$ pc at $z \sim 20$. Note that the gravitational forces due to
the baryonic core of the minihalo are not smoothed, since there is
high spatial resolution in this region. At their final timestep, all
simulations had more than 100 dark matter particles inside a proper
scale of 1~pc. 

The dark matter density profile is not resolved in the very central
regions of the halo where WIMP annihilation heating may be
important. Thus the results of these simulations are used as a guide
for two extrapolation methods (DM fits \#1 and \#2) to estimate the
dark matter density on scales of $\sim 10^{-5}-10^{-4}\:{\rm pc}$. We
fit power laws of the form
\begin{equation}
\rho_{\chi} = A_{\chi} \left(\frac{r}{\rm pc}\right)^{-\alpha_{\chi}}
\label{dmden}
\end{equation}
to the radial dark matter density profiles, evaluated over particular
ranges in radius. DM fit \#1, shown in Fig.~\ref{dm}, uses only data
points from the central regions with more than 100 particles per bin
(corresponding to scales from about 50~pc down to $r\gtrsim
1.5$~pc). We find $\alpha_{\chi}=1.35, 1.24, 1.65$ and $A_{\chi} =
1674, 1696, 2732 \:{\rm GeV\:cm^{-3}}$ for simulations A, B, C,
respectively. Note that simulation C is has a significantly steeper
density profile than A and B. All these density profiles are expected to
underestimate the dark matter density in the inner regions of the
halo, where the baryon density begins to dominate over dark matter and
adiabatic contraction of the dark matter tends to steepen its profile
(Blumenthal et al. 1986). Thus we consider DM fit \#1 to provide a
conservative lower limit to the dark mater density and thus the
heating rate.

Our second extrapolation method, DM fit \#2, is based on simulation
data from the inner (but not the innermost, $r<0.025$~pc) regions (see
Fig.~\ref{dm}).  For simulation A, we have $A_{\chi} = 2396\:{\rm
GeV\: cm^{-3}}$, $\alpha_{\chi} = 1.980$ when $r < 1$~pc (and
$A_{\chi} = 2409\:{\rm GeV\: cm^{-3}}$, $\alpha_{\chi} = 1.51$ when $r
> 1$ pc). For simulation B, $A_{\chi} = 1602\:{\rm GeV\: cm^{-3}}$,
$\alpha_{\chi} = 1.946$ when $r < 1$ pc (and $A_{\chi} = 1696\:{\rm
GeV\: cm^{-3}}$, $\alpha_{\chi} = 1.24$ when $r > 1$ pc). For
simulation C, we have $A_{\chi} = 942\:{\rm GeV\: cm^{-3}}$,
$\alpha_{\chi} = 2.108$ when $r < 0.1$ pc (and $A_{\chi} = 2732\:{\rm
GeV\: cm^{-3}}$, $\alpha_{\chi} = 1.645$ when $r > 0.1$ pc).  The
density profiles of the inner regions show a significant steepening
compared to the outer regions. This is likely to be the result of
adiabatic contraction, since it occurs at a scale $\sim 1$~pc where
the density of baryons begins to dominate. We emphasize again that the
simulation points on which DM fits \#2 are based correspond to bins
with only a small ($<100$) number of particles and must be treated
with caution. We do note, however, that on the very innermost scales
($\sim 0.01$~pc) the dark matter density profiles in the numerical
simulations are even steeper than the analytic DM fits \#2. 

If adiabatic contraction of dark matter in the baryon-dominated
potential is responsible for sculpting these profiles, one expects
that the maximum steepness of the dark matter density profile will be
equal to that of the baryons, which has been shown in simulations to
have an approximate power law profile of $\rho_{\rm gas}\propto
r^{-2.2}$ (see \S\ref{S:baryon}). The baryonic density profile
inevitably flattens in its center, so that the power law profile is
only valid inwards to some core radius, $r_c$. This core radius
shrinks as collapse proceeds. In situations where we are interested in
the global luminosity provided by WIMP annihilation in the halo
(\S\ref{S:equilibrium}), we will assume that the dark matter density
profile also exhibits this core radius, inside of which its density is
also constant.

\begin{figure}[!h]
\plotone{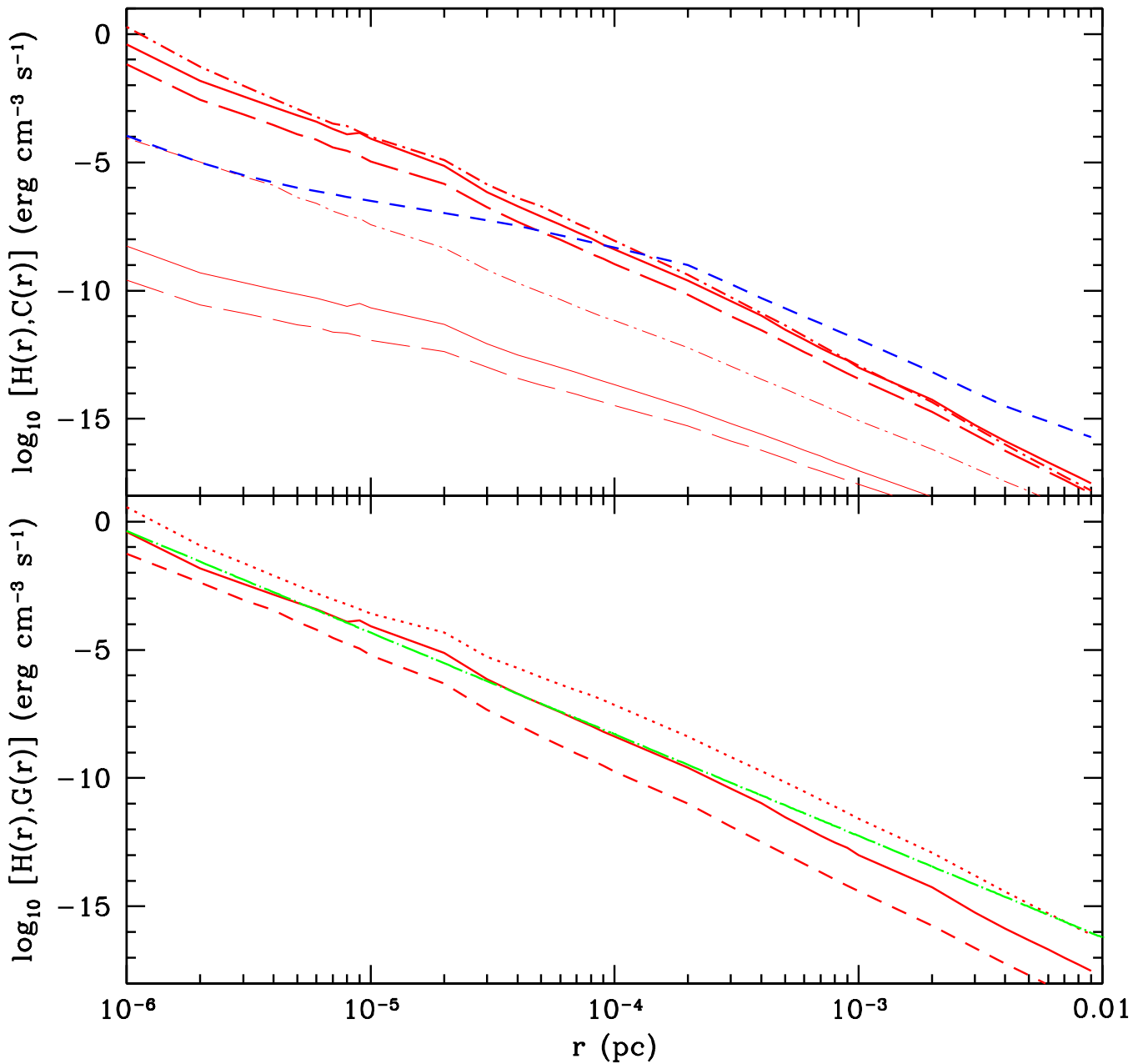}
\caption{Top panel: WIMP annihilation heating rate, $H(r)$, for
simulation A (solid lines), B (long-dashed lines) and C (dot-dashed
lines) with DM fit \#1 (lower, thin lines) and DM fit \#2 (upper,
thick lines). All these models assume $m_\chi=100$~GeV. The cooling
rate, $C(r)$, is shown by the dashed line (see
\S\ref{S:baryon}). Bottom panel: Variation of $H(r)$ with WIMP mass
for simulation A, DM profile \#2. WIMP masses $m_\chi$ = 10, 100 and
1000~GeV are shown by the dotted, solid, and dashed lines,
respectively. The dot-long-dashed line shows the WIMP annihilation
energy generation rate, $G(r)$, for $m_\chi=100$~GeV (it is to be
compared with the solid line). At large radii, $H(r)$ becomes much
smaller than $G(r)$, since not all the energy generated is
absorbed.\label{mass}}
\end{figure}

For the fiducial case, we assume a dark matter particle mass of
$m_\chi=100$~GeV, which is typical for a weakly interacting
particle. The photon multiplicity function $dN_\gamma/dE_\gamma$ for
the important annihilation channels is given by the form (Bergstr\"om
et al. 1998; Feng et al. 2001)
\begin{equation}
\frac{dN_\gamma}{dx} = \frac{a e^{-bx}}{x^{1.5}},
\label{photon_multiplicity}
\end{equation}
where $x = E_\gamma / m_\chi$ and $(a,b)$ are constants for the
particular annihilation channel. We use the values given in Feng,
Matchev \& Wilczek (2001), and average over the different annihilation
channels they considered to obtain $(a=0.9, b=9.56)$. We set $\left<
\sigma_a v \right > = 3 \times 10^{-26}\:{\rm cm^3\:s^{-1}}$
(e.g. Jungman et al. 1996). The heating rate as expressed in
equation~\ref{heating} was calculated for the above density profiles
and is shown in Fig.~\ref{mass}.

The variation of $H(r)$ on the WIMP mass, $m_\chi$, is also shown in
Fig.~\ref{mass}. We consider cases with $m_\chi=10$~GeV and 1~TeV,
i.e. factors of 10 below and above our fiducial value. Once the simple
$m^{-1}_\chi$ dependence of $H(r)$ (eq.~\ref{heating}) is accounted
for, we see that remaining variations in $H(r)$ are within about a
factor of 2. These are due to the $m_\chi$ dependencies of the photon
multiplicity function, the energy integral and the $ \left < S
(r,r',E) \right >$ function.

It is also informative to compare $H(r)$ to the energy generated by
WIMP annihilation per unit volume, per unit time
\begin{equation}
G(r) = \frac{\rho^2_{\chi}(r) \, \left< \sigma_a v \right >}{2 m_\chi} \int_0^1 dx \, x \, \frac{dN_\gamma}{dx}
\label{g}
\end{equation}
Fig. \ref{mass} compares $G(r)$ and $H(r)$ for Simulation A, for DM
fit \#2. At small distances ($r<100$ AU), the two curves are very
similar, whereas at larger distances, $G(r)\gg H(r)$ since not all the
energy generated is absorbed. It is also interesting to note that
$H(r)$ can exceed $G(r)$ at high densities, since unlike $G(r)$,
$H(r)$ is not a strictly local function of $r$.

\subsection{Baryon Density, Temperature, $\rm H_2$ Fraction and Cooling Rate}\label{S:baryon}

\begin{figure}[!h]
\plotone{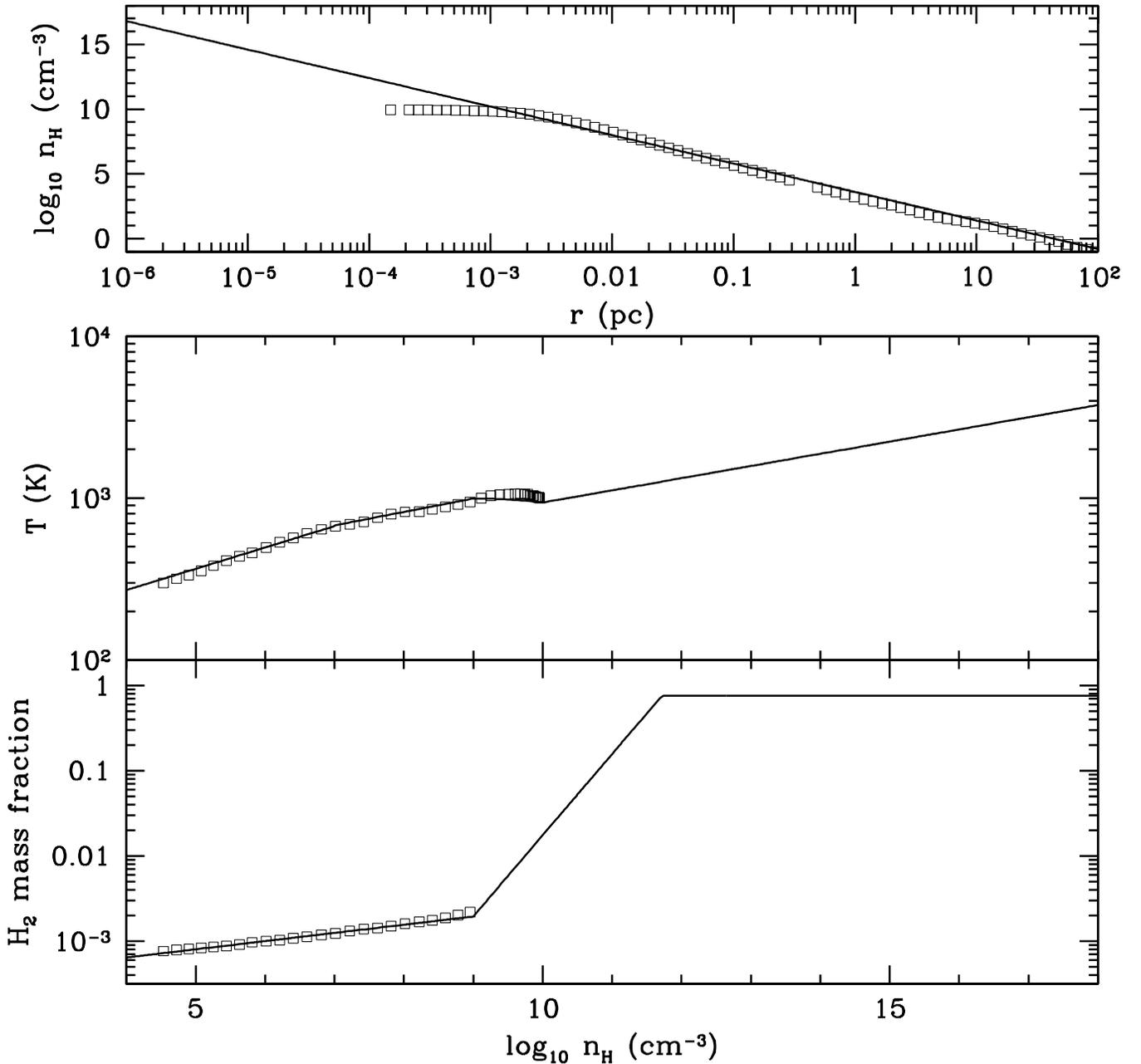}
\caption{Top panel: radially-averaged H number density, $n_{\rm H}$,
for simulation A (open squares). The solid line shows a power law
$n_{\rm H}\propto r^{-2.2}$, normalized from the simulations of
Yoshida et al. (2006) to have $\rm log_{10} (n_{\rm H}/{\rm cm^{-3}})
=13.15$ at $r=10~{\rm AU}=4.88\times 10^{-5}~{\rm pc}$. Middle panel:
radially-averaged temperature, $T$, versus radially-averaged H number
density, $n_{\rm H}$ for simulation A (open squares). The solid line
shows the analytic fits estimated from the results of Yoshida et
al. (2006). Bottom panel: radially-averaged $\rm H_2$ mass fraction
versus radially-averaged $n_{\rm H}$ for simulation A (open
squares). The solid line shows the analytic fits estimated from the
results of Yoshida et al. (2006).
\label{bar}}
\end{figure}

In the numerical simulations of O'Shea \& Norman (2007), collapse of
baryons was followed with adaptive mesh refinement to much higher
spatial resolution than the dark matter. Spatial scales down to $115
h^{-1}\rightarrow 164$~AU (comoving), i.e. about 10~AU (proper), were
resolved (see Fig.~\ref{bar}). However, in many cases we consider,
knowledge of the gas properties is required on even smaller
scales. Thus we utilize the results of the even higher resolution
smooth particle hydrodynamics (SPH) simulations of Yoshida et
al. (2006), which were carried out for a single minihalo in which the
collapse was followed to higher densities. The structure of baryonic
core is controlled mostly by the microphysics of $\rm H_2$ cooling, in
which case one expects a fairly universal profile of density,
temperature and chemical composition.

For the Hydrogen number density, $n_{\rm H}= 0.76 \rho_{\rm gas}/m_p$,
we use the analytic fit
\begin{equation}
n_{\rm H} = 1.41 \times 10^{13} \, \left( \frac{r}{10\:{\rm AU}}\right)^{-2.2}\:{\rm cm^{-3}},
\label{eq:rho_yoshida}
\end{equation}
which is based on conditions at 10~AU in the simulation of Yoshida et
al. (2006). Figure~\ref{bar} shows this fit is also a good description
of the gas in the simulations of O'Shea \& Norman (2007) at
$r\geq 200$~AU.
 
The temperature profile (as a function of density) at densities
$n_{\rm H}> 10^9$~cm$^{-3}$ was fit with an analytic function based on
the results of Yoshida et al. (2006). For lower densities, our
simulations A, B and C are reliable. We use the fitting functions (see
Fig.~\ref{bar}): $T = 165.5 \: (n_{\rm H}/{\rm cm^{-3}})^{0.0754}\:K$
for densities $n_{\rm H} > 10^{10}\:{\rm cm^{-3}}$; $\log (T/K) = 3 -
0.025 \log^2({n_{\rm H}}/10^9{\rm cm}^{-3})$ for densities $10^9\:{\rm
  cm^{-3}}< n_{\rm H}< 10^{10}\:{\rm cm^{-3}}$; $T = 178 (n_{\rm
  H}/{\rm cm^{-3}})^{0.083}\:K$ for densities $10^7\:{\rm cm^{-3}}<
n_{\rm H}< 10^{9}\:{\rm cm^{-3}}$; and $T = 80.5 (n_{\rm H}/{\rm
  cm^{-3}})^{0.1315}\:K$ for densities $n_{\rm H}<10^7\:{\rm cm^{-3}}$.

The molecular hydrogen mass fraction $f_{\rm mol}$ is also fit from
the results of the simulations of O'Shea \& Norman (2007) and Yoshida
et al. (2006). We use the fitting functions (see Fig.~\ref{bar}):
$f_{\rm mol} = 5 \times 10^{-12} (n_{\rm H}/{\rm cm^{-3}})^{0.954}$
for densities $n_{\rm H}>10^9\:{\rm cm^{-3}}$ (Yoshida et al. 2006)
and $f_{\rm mol} = 2.64 \times 10^{-4} (n_{\rm H}/{\rm
cm^{-3}})^{0.0963}$ for densities $n_{\rm H}<10^9\:{\rm cm^{-3}}$
(O'Shea \& Norman 2007).


We use the analytic results of Hollenbach \& McKee (1979) to
implement rotational and vibrational $\rm H_2$
cooling. Collision-induced emission (CIE) cooling follows the fit
published in Yoshida et al. (2006). We also account for the loss of
cooling efficiency due to the gas becoming optically thick.  The
opacity of the gas to cooling radiation was measured from the results
of Yoshida et al. (2006). The cooling efficiency, $\eta$, follows the
fitting function: $\eta = 9.4 \times 10^5 (n_{\rm H}/{\rm
cm^{-3}})^{-0.548}$ for densities $n_{\rm H} > 10^{13}\:{\rm
cm^{-3}}$; $\eta = 900 (n_{\rm H}/{\rm cm^{-3}})^{-0.316}$ for
densities $10^{11}\:{\rm cm^{-3}} < n_{\rm H} < 10^{13}\:{\rm
cm^{-3}}$; $\eta = 1.7 \times 10^5 (n_{\rm H}/{\rm cm^{-3}})^{-0.523}$
for densities $10^{10}\:{\rm cm^{-3}} < n_{\rm H} < 10^{11}\:{\rm
cm^{-3}}$; and $\eta=1$ for lower densities.

The cooling rate, $C(r)$, is shown in Fig.~\ref{mass} for simulation
A. The results for simulations B and C are very similar.

 
   

\subsection{Comparison of Heating and Cooling Rates and the Equilibrium Structure}\label{S:equilibrium}

\begin{figure}[!h]
\plotone{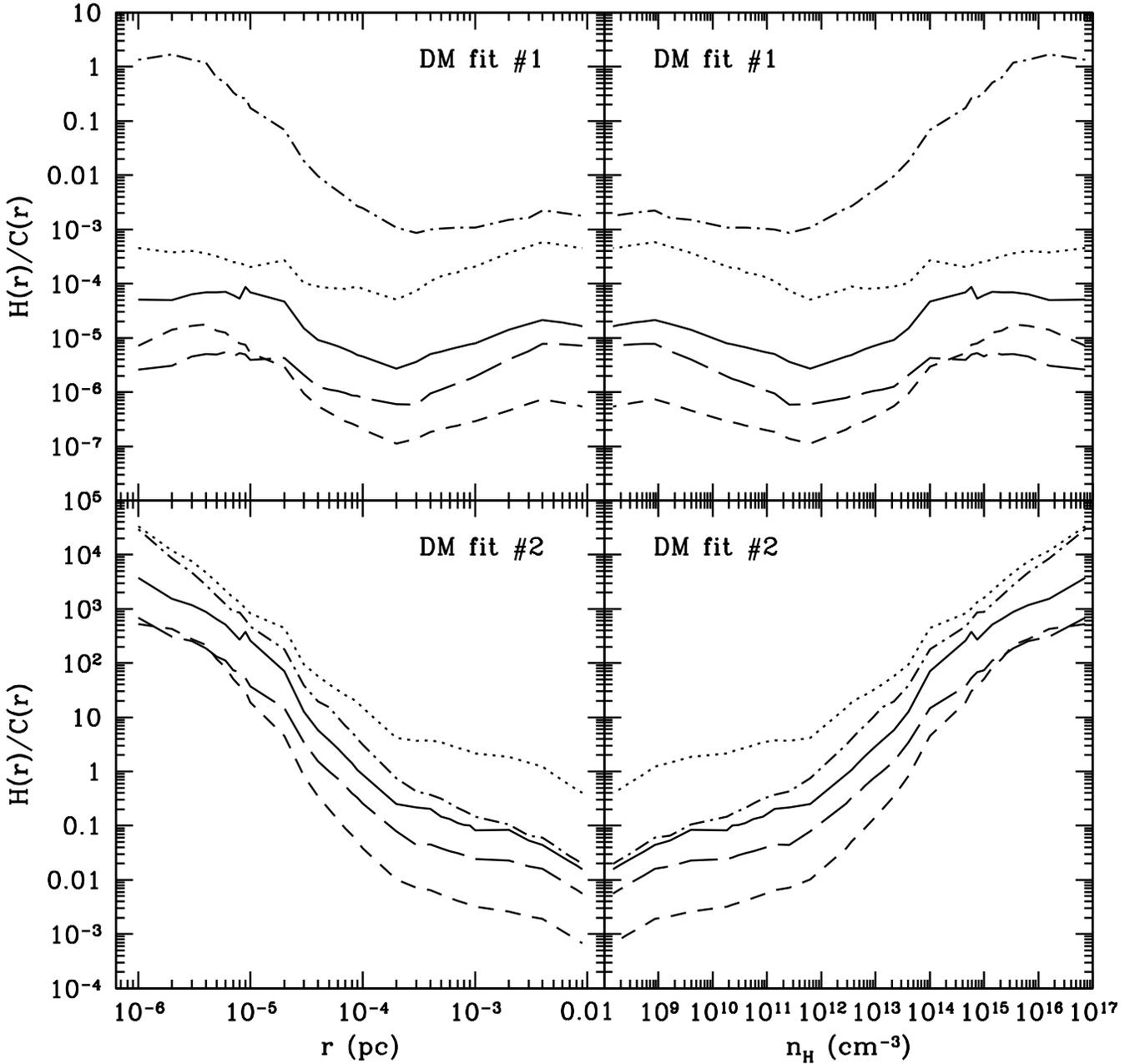}
\caption{Ratio of dark matter annihilation heating rate, $H(r)$, to
baryonic cooling rate, $C(r)$, for dark matter density fits \#1 (top
panels) and \#2 (bottom panels) for simulations A (solid lines), B
(long-dashed lines), C (dot-dashed lines) for
$m_\chi=100$~GeV. Results for simulation A with $m_\chi = 10$~GeV
(dotted lines) and $m_\chi = 1$~TeV (dashed lines) are also shown.
\label{hc}}
\end{figure}

Figure~\ref{hc} shows the radial profiles of the ratio of the heating
rate caused by WIMP annihilation to the gas cooling rate, $H(r)/C(r)$,
for simulations A, B, C and dark matter density fits \#1 and \#2,
assuming $m_\chi=100$~GeV. For simulation A, the dependence on
$m_\chi$ is also illustrated using $m_\chi=10$~GeV and
1000~GeV. Results are also shown as a function of $n_{\rm H}$, using
equation~\ref{eq:rho_yoshida} to derive the density at a given radius.

For the conservative, relatively shallow dark matter density profiles
(fit \#1) of simulation A and B and for a fiducial WIMP mass of
100~GeV, the heating rate is never larger than $\sim 10^{-4}$ of the
cooling rate for densities up to $n_{\rm H}=10^{17}\:{\rm
cm^{-3}}$. In this situation we expect that the star formation process
from these gas cores, in particular the protostellar structure and
accretion rate (e.g. Tan \& McKee 2004) would be unaffected by dark
matter annihilation heating. The expected mass of the stars forming
from these halos is then $\sim 100-200 M_\odot$, set by ionizing
feedback, especially disk photoevaporation, on the protostellar
accretion flow (McKee \& Tan 2008).

The case of DM fit \#1 for simulation C is somewhat steeper
($\alpha_\chi=1.65$) than for simulations A and B ($\alpha_\chi\simeq
1.3$), and this leads to the heating rate becoming comparable to the
cooling rate at $n_{\rm H}\gtrsim 10^{15}\:{\rm cm^{-3}}$. This trend
is continued for the steeper density profiles of DM fit \#2, all of
which have a central region where WIMP annihilation heating dominates
over cooling, typically for $r\lesssim 10^{-4}\:{\rm pc}$, i.e. 20~AU,
and $n_{\rm H}\gtrsim 10^{12}\:{\rm cm^{-3}}$. One important factor
setting this density scale is drop in the cooling efficiency of the
gas (Yoshida et al. 2006, their Fig.~4) because of the increasing
opacity of the gas.

These results are broadly consistent with the those of Spolyar et
al. (2008), showing that the properties of the dark matter heating
dominated core are relatively insensitive to the details of radiative
transport of the WIMP annihilation heating that we have included, and
that their analytic model for the dark matter density distribution is
similar to our DM fit \#2 to simulated minihalos (see also Freese et
al. 2008b).

The expected consequence of a central core where dark matter heating
of the gas dominates its radiative cooling is a halt to the
collapse. This process would then set the core radius, $r_c$, for both
the dark matter and baryon density distributions. To evaluate this
core radius and the total luminosity of the equilibrium structure, we
now consider the enclosed luminosity profile, $L(r)$, of the halos,
defined for WIMP annihilation heating as
\begin{equation}
L_\chi(r) = 4 \pi \int_0^r dr'  r'^2 \, H(r'),
\label{LDM}
\end{equation}
and for baryonic cooling as
\begin{equation}
L_{\rm gas}(r) = 4 \pi \int_0^r dr'  r'^2 \, C(r').
\label{LC}
\end{equation}
The luminosity due to WIMP annihilation will be dominated by the
central part of the halo for $\alpha_\chi>1.5$ (eq.~\ref{g}), which is
the case for all DM fits \#2. We solve for the core radius, $r_c$, for
each simulation via the condition $L_\chi(r\rightarrow\infty)=L_{\rm
gas}(r\rightarrow\infty)$, i.e.
\begin{equation}
\frac{4\pi r^3_c}{3} \, H(r_c) + 4\pi \int_{r_c}^\infty \, dr \, r^2 \, H(r) = \frac{4\pi r^3_c}{3} \, C(r_c) + 4 \pi \int_{r_c}^\infty \, dr \, r^2 \, C(r).
\end{equation}
This is an equilibrium condition in the sense that the energy
generation rate by dark matter annihilation heating balances that
radiated away by the baryons. Note that for the profiles analyzed
here, both heating and cooling luminosities are dominated by the
central regions, i.e. the integrals converge to a finite value as
$r\rightarrow\infty$. Note also that local heating will dominate local
cooling in an inner region that is somewhat larger than $r_c$, and
cooling will dominate heating outside of this region.

\begin{figure}[!h]
\plotone{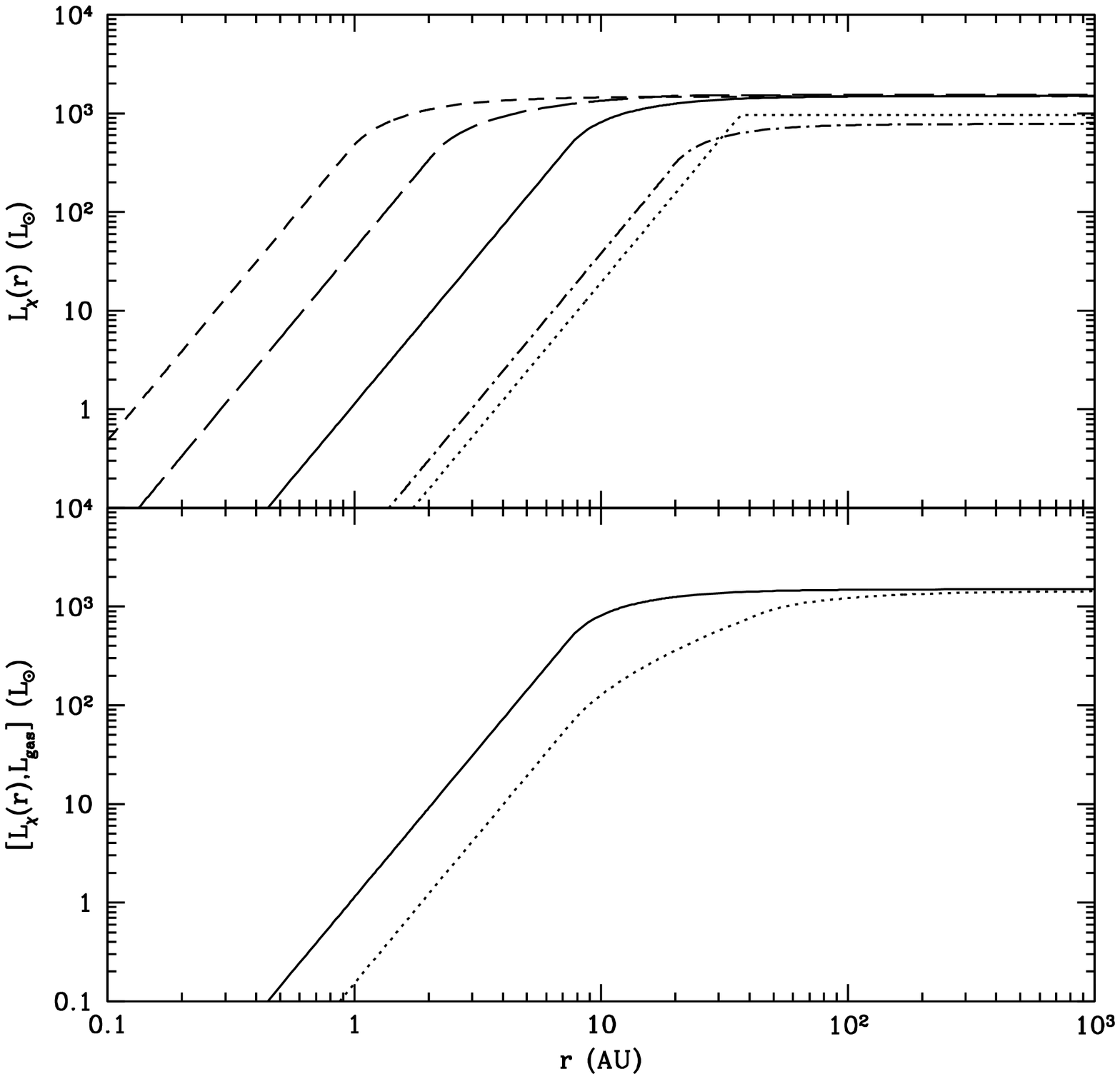}
\caption{Enclosed luminosity generated from the heating due to dark
matter annihilation, $L_\chi (r)$, for the equilibrium halo
configurations (see text). The top panel shows $L_\chi(r)$ for
simulations A (solid line), B (long-dashed line), C (dot-dashed line),
assuming $m_\chi = 100$~GeV and dark matter density fit \#2. The same
model for simulation A, but with $m_\chi = 10$~GeV is shown with the
dotted line and with $m_\chi = 1$~TeV is shown with the dashed
line. The bottom panel compares $L_\chi(r)$ for simulation A, dark
matter density fit \#2 and $m_\chi =100$~GeV (solid line) with the
baryonic cooling luminosity, $L_{\rm gas}(r)$, (dotted line) for the
same halo.
\label{lum}}
\end{figure}

Figure~\ref{lum} shows the luminosity profiles for DM fit \#2 for
simulations A, B and C with $m_\chi=100$~GeV, and for simulation A
with $m_\chi=10$~GeV and 1~TeV. Also shown is $L_{\rm gas}$ for the
fiducial model. With $m_\chi = 100$ GeV, the core radii for
simulations A, B, C are $r_c= 7.4, 2.2, 19$~AU, respectively, and
$L_\chi(r\rightarrow\infty) = 1460, 1540, 781 L_\odot$,
respectively. The dark matter density in these cores are $\rho_{\chi,c} =
1.5\times 10^{12}, 7.5\times10^{12}, 3.0\times 10^{11}\:{\rm
GeV\:cm^{-3}}$, respectively.  For simulation A with $m_\chi=10$~GeV and 1~TeV, we
find $r_c= 37, 0.93$~AU, respectively, $L_\chi(r\rightarrow\infty)
= 960, 1510 L_\odot$, respectively, and $\rho_{\chi,c} =
6.2\times 10^{10}, 9.2\times10^{13}\:{\rm
GeV\:cm^{-3}}$, respectively.

The constancy of these total luminosities is set by the baryonic
cooling luminosity, which is dominated by the region from $\sim 10$~AU
to $\sim 100$~AU, and is independent of the much smaller core radius,
which adjusts itself so as to provide enough WIMP annihilation
luminosity to match that radiated away by the baryons.

We note our solution has made the approximation that the baryonic
properties used to estimate $L_{\rm gas}$ do not include the effects
of WIMP annihilation heating. This effect would lead to hotter
temperatures and more efficient cooling, so that the actual
equilibrium structure would be somewhat smaller and denser.

Energy transport from the hotter inner part of the halo to the outer
cooler part, either via convection or radiation, would be needed to
set up a hydrodynamic equilibrium: i.e. a dark matter powered
(proto)star. The size of this star would be set by the $\tau=1$
surface, beyond which the energy flux from the interior is free to
escape. The results of Yoshida et al. (2006) imply that the cooling
efficiency has dropped to 0.1 by the time $n_{\rm H}=10^{12}\:{\rm
cm^{-3}}$, corresponding to about 30~AU. This is an approximate upper
limit to the initial size of the protostar, since the actual gas
temperature will be somewhat hotter and a larger fraction of the
radiated energy would be in the continuum (as opposed to line
radiation).

\subsection{Subsequent Evolution of the Dark Matter Powered Protostar}\label{S:protostar}

Since there is a large mass reservoir at large $r$ that is undergoing
cooling, baryons will continue to accrete to the central protostar,
deepening the gravitational potential and thus causing more dark
matter to be concentrated here also. For simulation A, DM fit\#2 and
$m_\chi=100$~GeV with $r_c=7.4$~AU, the mass inside $r_c$ is
0.17~$M_\odot$, and the accretion rate is close to $0.1\:M_\odot\:{\rm
yr^{-1}}$ with infall speeds of about 4~$\rm km\:s^{-1}$ (Yoshida et
al. 2006), which are mildly supersonic. As the central mass grows, the
infall is expected to become more supersonic, i.e. closer to the free
fall speed of the central protostellar mass at its surface.

The detailed evolution of this dark matter powered protostar is beyond
the scope of this paper. Freese et al. (2008c) have considered the
protostellar structure of such stars, but starting from a more evolved
stage when $m_*=3\:M_\odot$ and for $\dot{m}_*=2\times
10^{-3}\:M_\odot\:{\rm yr^{-1}}$. After 4,500~yr of accretion this
initial structure has grown in mass to 12~$M_\odot$ and has a radius
of 2.8~AU and a luminosity of $1.1\times 10^5\:L_\odot$ (ignoring
accretion luminosity). The dark matter in the star is assumed to
increase via adiabatic contraction of the surrounding halo and enough
luminosity is provided by dark matter annihilation to support stars of
at least 1000~$M_\odot$, for which $L_*=4\times 10^6\:L_\odot$. Being
much larger and cooler than protostars on the zero age main sequence (ZAMS),
these dark matter powered protostars are expected to have relatively
weak radiative feedback on their accretion envelopes and thus may be
able to attain very high masses.

An alternative possibility is that the baryonic mass of the star grows
more rapidly than the enclosed dark matter mass and that this more
massive star is unable to be supported by WIMP annihilation
luminosity.  The star contracts and reaches central densities and
temperatures at which nuclear fusion of H starts. Note that fiducial
models of Population III.1 protostars (e.g. Omukai \& Palla 2003; Tan
\& McKee 2004) do not expect contraction to the main sequence until
$m_*\simeq100\:M_\odot$, i.e. when the Kelvin-Helmholz time becomes
less than the accretion growth time. The critical feedback process,
disk photoevaporation, theorized to limit the masses of Pop III.1
stars is expected to truncate protostellar growth along the ZAMS at
about $140\:M_\odot$, depending on the accretion rate (McKee \& Tan
2008). Thus if WIMP annihilation were able to support the protostar
only up to $\sim 100\:M_\odot$, there would be only modest
implications for the initial mass function of the first stars.

Understanding the accumulation of dark matter by the protostar is of
critical importance for deciding between these two scenarios.  There
are processes and effects that can cause inefficient adiabatic
contraction of the dark matter halo in response to protostellar
accretion. Rapid free-fall collapse of the baryons joining the
star-disk system is expected after the expansion wave moves out.
Standard analytic analyses of adiabatic contraction typically assume
very slow changes in the baryonic potential (e.g. Blumenthal et
al. 1986). For rapid collapse, one expects a delayed response of the
dark matter, and during this time $m_{\rm *DM}/m_*$ will have
decreased. 

Another effect is excitation of the dark matter velocity
dispersion by scattering with baryonic density fluctuations. These
will probably occur most strongly in the accretion disk around the
protostar, since such fluctuations are required for angular momentum
transport, and, if not initially present, will inevitably develop via
gravitational instability as the disk mass grows (Tan \& Blackman
2004).  For accretion from a core with a given density and angular
momentum, the disk size can be described via (Tan \& McKee 2004):
\begin{equation}
r_d = 3.44 \left(\frac{f_{\rm Kep}}{0.5}\right)^2 \left(\frac{m_{*d}}{M_\odot}\right)^{9/7} K'^{-10/7}\:{\rm AU},
\end{equation}
where $f_{\rm Kep}$ is the ratio of the rotational to Keplerian
velocities of the gas at the sonic point of the infall and has a
fiducial value of 0.5 in the simulations of Abel et al. (2002),
$m_{*d}$ is the mass of central protostar and its disk (we typically
assume a disk mass that is one third of the protostellar mass), and
$K'$ is the normalized entropy parameter (related to the density of
the core at the time of protostar formation: a denser core has a
larger value of $K'$), with fiducial value of 1.0. We see that the
expected disk sizes when the central mass is $1, 10, 100~M_\odot$ are
$3.4, 66, 1300$~AU. Most mass will join the dark matter powered
protostar via a disk once $m_{*d}\gtrsim$~a few solar masses.

If the initial protostar fails to accumulate any addition dark matter and
retains what it has during Kelvin-Helmholz contraction (which is
expected since the contraction time is much longer than the dynamical
time), then, ignoring depletion (discussed below, but which occurs on
timescales much longer than the initial accretion growth time), the
WIMP annihilation luminosity of the star is approximately described by
\begin{equation}
L_\chi \simeq L_{\chi,0} \left(\frac{r_*}{r_{*,0}}\right)^{-3},
\end{equation}
where $L_{\chi,0}\simeq 10^3\:L_\odot$ and $r_{*,0}\simeq 10$~AU are
fiducial values for the initial luminosity and radius of the star,
respectively, and assuming homologous evolution of the density
profile. The protostellar size in models with $L_\chi=0$ is expected
to be $\sim 1$~AU for $m_*\lesssim 10\:M_\odot$ (Tan \& McKee
2004). Compressing a protostar to this size would lead to
$L_\chi\simeq 10^6\:L_\odot$ for the above starting conditions, which
is enough luminosity to support a $\sim 100\:M_\odot$ star. So in this
case the protostar would be able to be supported by WIMP annihilation
giving it $\gtrsim$AU sizes until about $100\:M_\odot$. The ZAMS
radius of a $100\:M_\odot$ star is about $4\:R_\odot = 0.02$~AU, so it
is clear that contraction to this scale would not be possible until
very large stellar masses, unless the WIMP content of the protostar
became depleted.

These conclusions are of course sensitive to the initial condition. We
expect $L_{\chi,0}\simeq 10^3\:L_\odot$ to be relatively robust as it
is set by the baryonic cooling properties of the halo, but the initial
size of the dark matter core that provides this WIMP annihilation
luminosity depends quite sensitively on the uncertain WIMP mass and
the spatial distribution of the dark matter. For example, if the dark
matter core has an initial size of about 1~AU (e.g. similar to
simulation B, DM fit\#2, $m_\chi=100$~GeV and models with
$m_\chi=1$~TeV) and assuming the initial protostellar size scales
proportionately and is also about equal to this size, then the potential
luminosity available from contraction of the protostar to a given
radius would be a thousand times less than the case considered
above. Nevertheless, a contraction to $10\:R_\odot$ from 1~AU, would
still lead to a WIMP annihilation luminosity of $8\times
10^6\:L_\odot$.

We now compare the growth time of the protostar (i.e. the time since its formation, its age),\\ $t_*=2.92
\times 10^4 K'^{-15/7}(m_*/100M_\odot)^{10/7} \: {\rm yr}$ (Tan \&
McKee 2004) to the WIMP depletion time:
\begin{equation}
t_{\rm dep} = \frac{\rho_\chi}{\dot{\rho}_\chi} \simeq \frac{m_\chi}{\rho_\chi <\sigma_a
v>} \rightarrow 105 \frac{m_\chi}{100~{\rm GeV}}\left(\frac{\rho_\chi}{10^{12}{\rm GeV\:cm^{-3}}}\right)^{-1}\:{\rm Myr}.
\end{equation}
If the protostar contracts from an initial radius of 10~AU (where we
find $\rho_\chi\simeq 10^{12}\:{\rm GeV\:cm^{-3}}$,see above) to 1~AU
then $t_{\rm dep}\simeq 10^5\:{\rm yr}$. We see that, if replenishment
of WIMPs in the protostar is negligible, then depletion can become
important for AU scale protostars of $\sim 100\:M_\odot$. These
timescales are sensitive to $m_\chi$ and the dark matter density in
the initial equilibrium core, which shows significant variation between
simulation A, B, and C.

\section{Conclusions}\label{S:conclusions}

We have investigated the effects of WIMP dark matter annihilation on
the formation of Population III.1 stars by analyzing the results of
cosmological simulations that follow the gravitational collapse of
baryons and dark matter. While these simulations (O'Shea \& Norman
2007; Yoshida et al. 2006) have followed the baryons to very high
densities at scales $\lesssim 1$~AU, the dark matter is only
well-resolved down to scales $\sim 1$~pc. Thus we have considered
various power law extrapolations of the dark matter density profile
towards the center.

If one assumes the dark matter profile is self-similar, extending
inwards from the dark matter dominated regime with $\rho_{\chi}\propto
r^{-\alpha_{\chi}}$ and $\alpha_{\chi}\simeq 1.5$, then, for a
fiducial WIMP mass of 100~GeV, the dark matter annihilation heating is
typically negligible ($\sim10^{-4}$ of the cooling rate). This conclusion
would be unchanged for a reduction in the WIMP mass by a factor of 10
or more. One of the simulated minihalos (C) exhibits a slightly
steeper density profile in the well-resolved region
($\alpha_\chi=1.65$, and in this case WIMP annihilation heating does
become important at $n_{\rm H}\gtrsim 10^{15}\:{\rm cm^{-3}}$.

However, there are theoretical reasons to expect that the dark matter
density profile will steepen because of adiabatic contraction in the
baryon-dominated core. Indeed, this process appears to be occurring in
the simulations of O'Shea \& Norman (2007), although it is not
well-resolved. A value of $\alpha_{\chi}\simeq2.0$ appears to be a
better description of the dark matter density profile in this
region. For such a profile, and again for a 100~GeV particle, the dark
matter annihilation heating now exceeds baryonic cooling for densities
$n_{\rm H}>10^{14}\:{\rm cm^{-3}}$, in agreement with the previous
study of Spolyar et al. (2008).

We considered the properties of equilibrium halos in which the density
distributions of the baryons and dark matter exhibit a constant
density central core. The luminosity that is generated is $\sim
10^3\:L_\odot$ and is relatively invariant, being set by baryonic
cooling processes. The sizes of the central cores range from $\sim 1$
to 40~AU.

The detailed effects of this extra heating on the protostellar
structure remain to be determined. We expect that subsequent baryonic
growth of the protostar will occur more rapidly than its accumulation
of dark matter, because the baryons are undergoing rapid free fall
collapse followed by disk accretion. However, even if the initial
protostar does not gain any additional dark matter, its initial dark
matter content in the subsolar mass core could be sufficient to
prevent contraction to the zero age main sequence for masses of
100~$M_\odot$ or greater. Such circumstances could have dramatic
implications for the masses of Pop III.1 stars, conceivably
raising the mass scale to a regime important for the formation of
supermassive black holes. These conclusions depend sensitively on the
initial protostellar core, which now needs to be studied with
self-consistent cosmological simulations that include the influence
of WIMP annihilation heating on the baryons.


\acknowledgements 

We thank T. Abel, K. Freese, C. McKee, D. Spolyar, and N. Yoshida for
discussions. The research of JCT is supported by NSF CAREER grant
AST-0645412. A.N. acknowledges financial support from the Deutsche
Forschungsgemeinschaft(DFG) International Research Training Group GRK
881.




\end{document}